\documentclass[aps,showpacs,superscriptaddress,altaffilletter]{revtex4-1}

\usepackage{color}
\usepackage{amssymb}
\usepackage{graphicx}
\usepackage[nodots,nocompress]{numcompress}
\usepackage{subfloat}

\begin{document}

\title{Evidence for $\nu_\mu \to \nu_\tau$ appearance in the CNGS neutrino beam\\ with the OPERA experiment}
                        
\author{ N.~Agafonova }  \affiliation{ INR Institute for Nuclear Research, Russian Academy of Sciences RUS-117312, Moscow, Russia }               
\author{ A.~Aleksandrov } \affiliation{ INFN Sezione di Napoli, I-80125 Napoli, Italy } 
\author{ A.~Anokhina } \affiliation{ SINP MSU-Skobeltsyn Institute of Nuclear Physics, Lomonosov Moscow State University, \mbox{ RUS-119992 } Moscow, Russia }              
\author{ S.~Aoki } \affiliation{ Kobe University, J-657-8501 Kobe, Japan }                             
\author{ A.~Ariga } \affiliation{ Albert Einstein Center for Fundamental Physics, Laboratory for High Energy Physics (LHEP), University of Bern, CH-3012 Bern, Switzerland }           
\author{ T.~Ariga } \affiliation{ Albert Einstein Center for Fundamental Physics, Laboratory for High Energy Physics (LHEP), University of Bern, CH-3012 Bern, Switzerland }   
\author{ T.~Asada } \affiliation{ Nagoya University, J-464-8602 Nagoya, Japan }    
\author{ D.~Autiero } \affiliation{ IPNL, Universit\'e Claude Bernard Lyon 1, CNRS/IN2P3, F-69622 Villeurbanne, France }  
\author{ A.~Ben~Dhahbi} \affiliation{ Albert Einstein Center for Fundamental Physics, Laboratory for High Energy Physics (LHEP), University of Bern, CH-3012 Bern, Switzerland }                                   
\author{ A.~Badertscher } \affiliation{ ETH Zurich, Institute for Particle Physics, CH-8093 Zurich, Switzerland }                    
% ...

\collaboration{OPERA Collaboration}
\date{\today}
             
\begin{abstract}
The OPERA experiment is designed to search for $\nu_{\mu} \rightarrow
\nu_{\tau}$ oscillations in appearance mode i.e. through the direct
observation of the $\tau$ lepton in $\nu_{\tau}$ charged current
interactions.
The experiment has taken data for five years, since 2008, with
the CERN Neutrino to Gran Sasso beam. 
Previously, two $\nu_{\tau}$ candidates with a $\tau$ decaying into hadrons were observed
in a sub-sample of data of the 2008-2011 runs.
Here we report the observation of a third $\nu_\tau$
candidate in the $\tau^-\to\mu^-$ decay channel coming from 
the analysis of a sub-sample of the 2012 run.  
Taking into account the estimated background,
the absence of $\nu_{\mu} \rightarrow \nu_{\tau}$
oscillations is excluded at the {\color{black}{3.4}}~$\sigma$ level.
\end{abstract}


\maketitle

\section{Introduction}
The OPERA experiment is designed to perform a crucial test of
neutrino oscillations aiming at the direct observation of
the appearance of $\tau$ neutrinos in a $\nu_\mu$ beam.
Using atmospheric neutrinos, the Super-Kamiokande experiment recently reported the evidence
for a $\nu_\tau$  appearance signal on a statistical basis and
with a low signal-to-noise ratio \cite{art14}.
The OPERA apparatus has the capability of detecting the $\nu_\tau$ charged-current
interactions on an event-by-event basis and with an extremely high signal-to-noise ratio. 
A positive evidence from OPERA can prove that the $\nu_\mu \to \nu_\tau$ transition
is the mechanism underlying the disappearance of muon
neutrinos at the atmospheric scale \cite{art13}, thus providing essential
support to the establishment of the 3-flavour mixing scheme.

To accomplish this task,
several ingredients are required: a high-energy neutrino beam, a long
baseline and a \mbox{kt-scale} detector with sub-micrometric
resolution. The CERN Neutrinos to Gran Sasso (CNGS \cite{ref:CNGS2}) beam was designed to deliver
muon neutrinos with a mean energy of 17 GeV to the Gran Sasso underground laboratory (LNGS)
where the detector is installed
at a distance of 730~km.
The contaminations of $\bar{\nu}_\mu$ and $\nu_e + \bar{\nu}_e$ charged current interactions at LNGS, relative to the number of $\nu_\mu$ charged current interactions,
are respectively 2.1\% and 0.9\%. The contamination from prompt  $\nu_{\tau}$ 
 is negligible. 

The OPERA detector \cite{OPERAdet} is composed of two identical
supermodules, each consisting of an iron spectrometer downstream of a
target section.  The target has a mass of about 1.2 kt and a
modular structure with approximately 150000 target units, called bricks. 
A brick is made of 56 1 mm-thick lead plates, acting as targets,
interleaved with 57 nuclear emulsion films, used as micrometric
tracking devices.  Each film is composed of two 44 $\mu$m-thick
emulsion layers on both sides of a 205 $\mu$m-thick plastic base.
Bricks are arranged in walls interleaved with
planes of scintillator strips forming the Target Tracker (TT).
The magnetic spectrometers, which consist of iron magnets
instrumented with Resistive Plate Chambers (RPC) and Precision
Drift Tubes (PT), are 
used for the measurement of the muon charge and momentum. 
For each event, the information provided by the electronic 
detectors allows assigning to each brick a probability to contain 
the neutrino interaction vertex.

In a recent publication \cite{secondtau} the $\nu_\tau$ appearance 
analysis
was described in detail, explaining 
the selection of  signal candidate events and the assessment of efficiencies and backgrounds.
In the following, the experimental procedure is briefly summarised along with
the description of a third $\nu_\tau$ candidate.

\section{Data Sample}
The CNGS beam has run for five years, from 2008 till the end of 
2012, delivering a total of $17.97 \times 10^{19}$ protons on
target yielding 19505 neutrino interactions recorded
in the targets.

A neutrino interaction is classified as being either charged current-like (1$\mu$) or
neutral current-like (0$\mu$) by using the electronic data of the TT and
the spectrometers.  The neutrino interaction vertex brick 
predicted by
the 
electronic detector data is extracted from the
target by an automatic brick manipulator system.
If the scanning of a dedicated pair of emulsion films (Changeable Sheets, CS), acting as 
an interface between the brick and the TT, yields tracks related to the neutrino interaction, the emulsion films of the brick are developed and distributed to the different scanning laboratories 
of the collaboration. Their analysis provides the
 three-dimensional reconstruction of the neutrino interaction and
 of possible secondary decay vertices of short-lived particles with micrometric accuracy.

The sample of events considered in our previous publication
\cite{secondtau} consisted of:
\begin{itemize}
\item all the $0\mu$ events collected and searched for in the two most probable bricks for the 2008-2009 runs and in the most probable one for the 2010-2011 runs,
\item all the $1\mu$ events with $p_\mu<15$~GeV/$c$ collected and searched for in the two most probable bricks for the 2008-2009 runs and in the most probable one for the 2010 run. 
\end{itemize}
  
Two $\nu_\tau$ candidate events in the hadronic decay
channels were observed: the first in the 2009 run data with a
one-prong topology~\cite{nutau}, the second in the 2011 run data with a 
three-prong topology~\cite{secondtau}.

The analysis was then performed on the
most probable bricks of $1\mu$ events with $p_\mu<15$~GeV/$c$
collected during the 2011 and 2012 runs. 
A $\nu_\tau$ candidate event in the muonic decay channel was
observed in this data sample. 

\section{The new $\nu_\tau$ candidate event}
\label{sect:pq}
\begin{figure}
\centering
\begin{center}
\includegraphics[width=14cm,height=13cm]{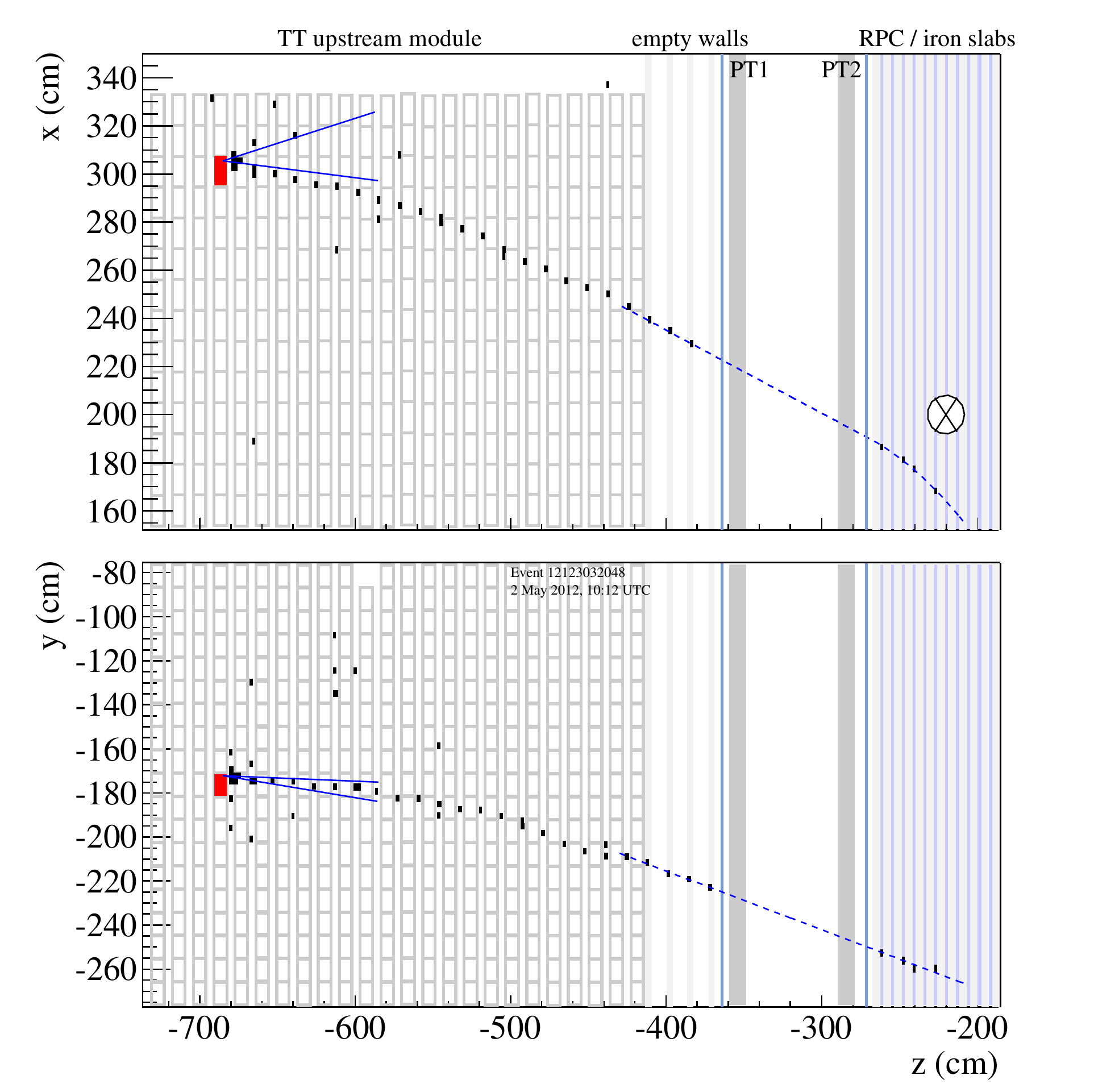}
\end{center}
\caption{Electronic detector display of the new $\nu_\tau$ candidate event.   
The blue solid lines represent the linear extrapolation of tracks measured in the 
emulsion films of the vertex brick.  The dashed blue lines show the fit of the most downstream hits according to the model:
\mbox{$x(z) = p_{0x} + p_{1x} (z-z_0 ) + p_{2x}  (z-z_0 )^2$} with $z_0 = -267.826$ cm. 
The quadratic term parameter is $p_{2x} = (-0.00389 \pm 0.00069)$~cm$^{-1}$ 
and the  fit $\chi^2/ndf$ is $2.6/4$.
\label{fig:fig1}}
\end{figure}

Figure \ref{fig:fig1} shows the electronic
detector display of this new event; the neutrino vertex brick (highlighted) 
is well contained in the target region.
An isolated, penetrating track is reconstructed in the electronic
detectors: 
the particle is recorded in 24 planes
of the TT and crosses 6 RPC planes before stopping in the
spectrometer.  This range corresponds to 1650~g/cm$^2$ of
material.

Four RPC planes have hits associated to the track in both
projections (Fig. \ref{fig:fig1}). The time spread of the RPC hits is within 20 ns.  The efficiency of the RPC planes was monitored
with cosmic ray tracks and muons from neutrino interactions in the
rock \cite{operanote}.

The muon momentum at the interaction vertex is accurately estimated
from the range of the particle in the electronic detector: a
Kalman filter-based algorithm yields a value of $p_\mu = (2.8 \pm
0.2)$~GeV/$c$.  The momentum estimate resulting from a measurement
of Multiple Coulomb Scattering (MCS)~\cite{mcs} in the downstream
brick, based on the emulsion data, leads to a compatible value
of $3.1^{+0.9}_{-0.5}$ GeV/$c$.

For events to be retained as candidates in the $\tau\to\mu$ decay channel the charge of the secondary muon track must either be measured to be negative or be undetermined. This requirement is applied in order to minimize the background from charged current $\nu_\mu$  interactions with production of a charmed particle decaying to a positive muon and where the primary negative muon goes undetected.

The charge measurement is performed using the bending of the track
in the magnetised iron given by the four available RPC hits.

For this event, no hits could be recorded by the 
PT planes (grey rectangles in Fig.~\ref{fig:fig1}) due to an inefficiency of the trigger (the trigger is given by a 2 out of 3 majority of the first RPC
plane within the magnet and two upstream dedicated RPC planes, the
so-called XPC planes \cite{OPERAdet}). 

As the last brick-filled wall is followed by three double layers of
TT planes (and essentially no other material, see Fig.~\ref{fig:fig1}),
the slope of the track at the entrance of the spectrometer could be
determined with the needed precision.

The relative alignment of RPC and TT planes in the $x$ and $y$ directions
was determined using muons from interactions of
neutrinos in the rock (horizontal tracks) and cosmic ray muons
(sensitive also to vertical displacements). In both cases,
the resulting alignment is accurate at the mm level.
The errors on the measurement points were computed assuming uniform
probability density across the strips that have a width of 2.6~cm ($x$) 
and 3.5~cm ($y$) for RPC detectors and 2.64~cm for the TT. 
The uncertainty related to MCS has also been taken into account.

The TT and RPC hits are fitted with a simple analytical model consisting
of a straight line in the 
field-free region matched to a parabola
in the magnetised region (Fig.~\ref{fig:fig1}).  
The fitted parabola
bends towards smaller $x$
(see Fig.~\ref{fig:fig1} top)
corresponding to a negative charge. The quadratic parameter is nonzero 
at 5.6~$\sigma$ significance.  The associated momentum at the
spectrometer entrance is compatible with the one measured from range.

\begin{figure}
\centering
\begin{center}
\includegraphics[width=10cm,height=5.5cm]{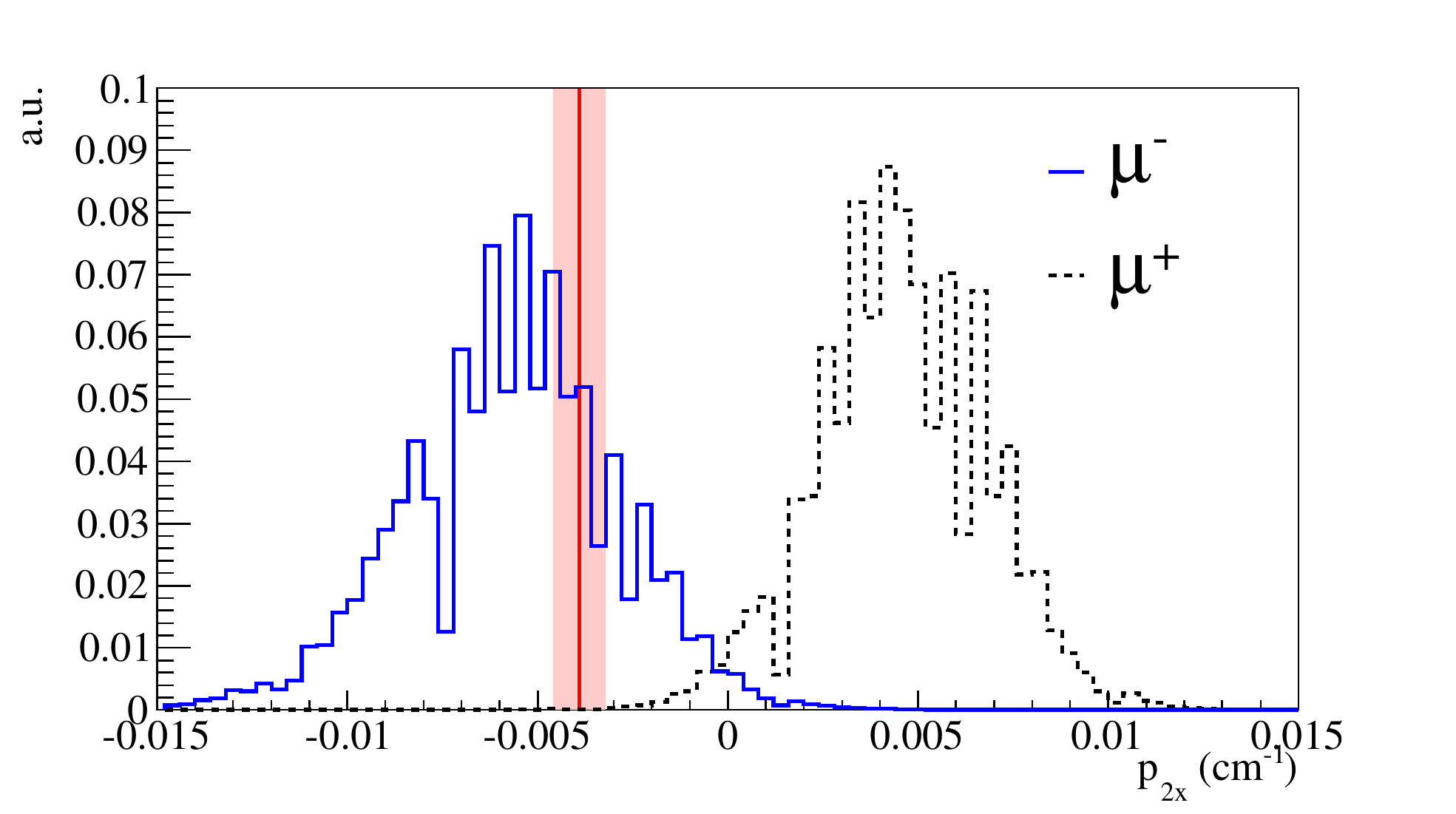}
\end{center}
\caption{ Distributions of $p_{2x}$ for a Monte Carlo sample of
  $\mu^-$ and $\mu^+$. 
The vertical solid line represents the measured value and the vertical
band corresponds to the 1 $\sigma$ confidence interval. The
visible structures are due to the pitch of the RPC readout which
introduces a discretisation effect.
\label{fig:fig1a}}
\end{figure}

For this event, the charge misidentification probability was estimated by means of a Monte Carlo simulation.
Muons of either charge were sampled
using a uniform momentum distribution.
A gaussian smearing of the incoming direction was
applied to account for the measurement error in the three most downstream TT
layers.  Only those muons stopping in the same slab observed for the event and with hits within a time window of 30~ns 
were selected, giving a sample of negative (positive) muons with an average momentum of 613
MeV/$c$ with a 48 MeV/$c$ r.m.s. (540 MeV/$c$ with a 26 MeV/$c$ r.m.s.).  
The strip read-out was simulated using the efficiency observed in real data.
The resulting distributions of the quadratic term $p_{2x}$ for the $\mu^-$ and 
$\mu^+$ samples are shown in
Fig.~\ref{fig:fig1a}.
The fraction of $\mu^+$ for which the bending is reconstructed as negative and that mimics a $\mu^-$ is 2.5\%, but for only 0.063\% of the $\mu^+$ the bending is more negative than the observed one.
The measured $p_{2x}$ is compatible with the
peak value of the distribution for the $\mu^-$ sample. For this event, the sign of the charge of the secondary muon is thus univocally determined to be negative. The efficiency of the RPC chamber downstream of the iron slab layer
in which the muon is assumed to have stopped is 93\%. If the muon had actually
stopped in the next layer with this RPC plane being inefficient,
then its momentum would have been larger by only 75 MeV/$c$, the precision in
its charge determination remaining unaffected.

\section{Event topology and kinematics}
The scanning of the CS films of the interaction brick yielded a track
matching the muon direction. A converging 
 pattern of more tracks was also found, reinforcing the
conditions for event validation.

\begin{figure}
\centering
\begin{center}
\includegraphics[width=12cm,height=10cm]{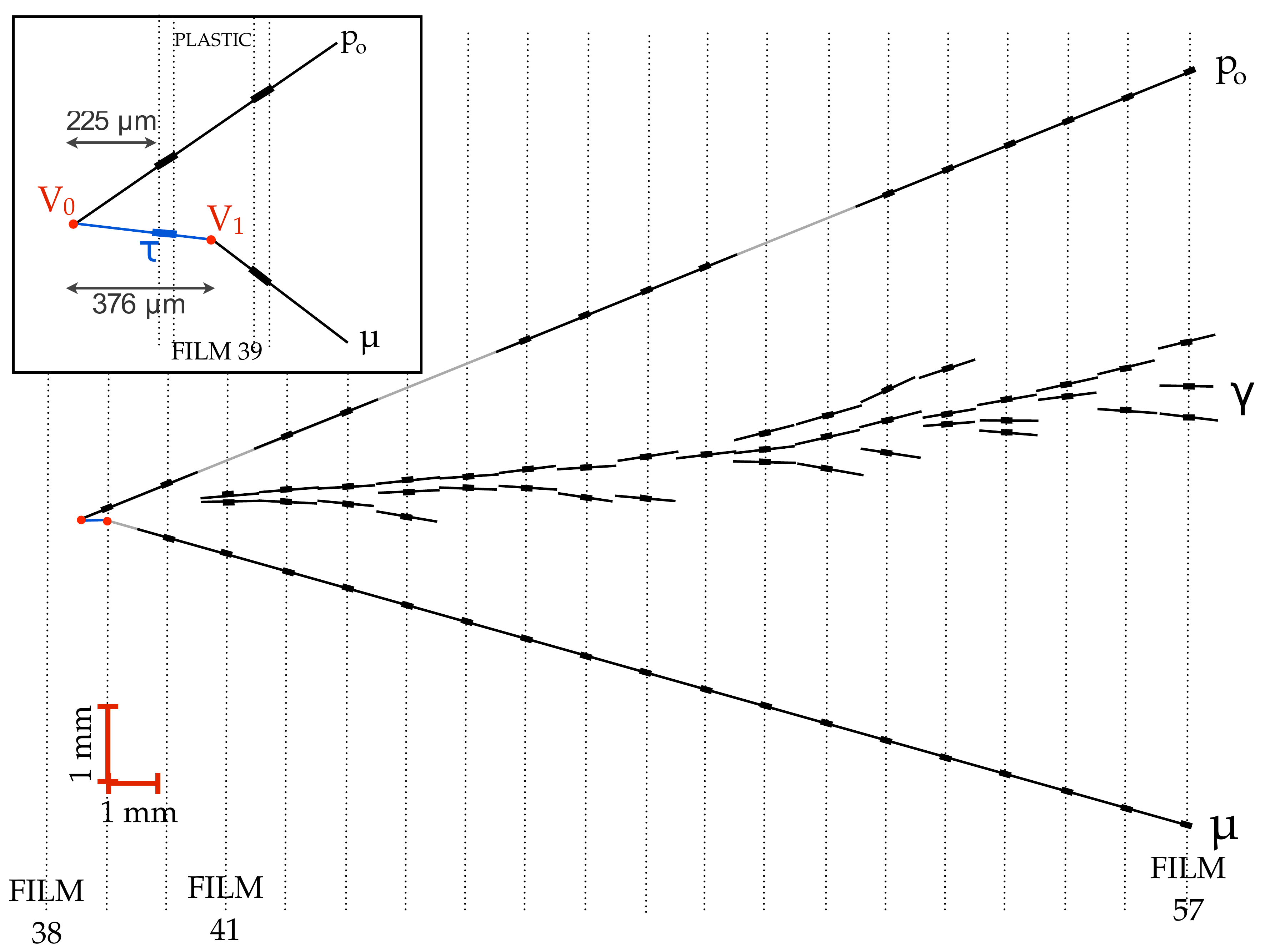}
\end{center}
\caption{Display of the new $\nu_\tau$ candidate event in the $xz$
  projection: tracks $\tau$ and $p_0$ come from the primary vertex;
  the $\tau$ candidate decays in the plastic base of film 39, track
  $d_1$ is the $\tau$ decay daughter identified as a muon. The
  starting point of the shower generated from the $\gamma$ is visible
  in film 41. The inset contains a zoomed view of the primary and decay vertex region.
\label{fig:fig6}}
\end{figure}

The neutrino interaction occurred well inside the brick with respect
to the longitudinal direction, 3.3~$X_0$ from its downstream face.
All tracks possibly related to the interaction were searched for in the brick with an angular acceptance
up to $\tan \theta = 1$.  The display of the event as reconstructed in
the brick is shown in Fig.~\ref{fig:fig6}.

The primary vertex ($V_0$) is given by two tracks: the $\tau$ lepton
candidate and a hadron track ($p_0$) having a distance of closest approach of
\mbox{$(0.5 \pm 0.5)$} $\mu$m. An electromagnetic shower produced by a
$\gamma$-ray and pointing to this primary vertex has also been observed.  The $\tau$ lepton decay occurs in the
plastic base of the film immediately downstream of the primary
vertex, after a flight length of \mbox{($376 \pm 10$)} $\mu$m. The
longitudinal coordinate of the decay vertex ($V_1$) with respect to the downstream face
of the lead plate containing the primary vertex ($z_{dec}$) is \mbox{$(151
\pm 10)$} $\mu$m. 

The kinematical quantities of the tracks measured in the emulsion films
are given in the following:
\begin{itemize}  
\item track $p_0$ has a momentum $p_{p_0}=(0.90^{+0.18}_{-0.13})$
  GeV/$c$, measured by MCS. It was found in the CS
  films. It was followed
  into the downstream brick where it disappears after having crossed
  18 lead plates. No charged particle track could be detected at the
  interaction point. It is classified as a hadron by its
  momentum-range correlation \cite{secondtau};

\item track $d_1$ is the $\tau$ decay daughter. Its angle with the
  $\tau$ lepton track ($\theta_{kink}$) is \mbox{$(245 \pm 5)$
  mrad}.  The impact parameter with respect to the primary vertex is
  \mbox{$(93.7 \pm 1.1)$ $\mu$m}. The track, found also on the CS films, 
  agrees with the muon track reconstructed in the electronic
  detectors in both momentum (\mbox{$\Delta
    p=0.3^{+0.9}_{-0.5}$~GeV/$c$}) and angle (\mbox{$\Delta
    \theta = 18 \pm 25$ mrad});

\item the shower originating from a $\gamma$-ray conversion has an
 energy of $(3.1^{+0.9}_{-0.6})$~GeV. The
  conversion to an $e^+e^-$ pair is observed 2.1~mm (0.36~$X_0$)
  downstream of the primary vertex to which it points
with an impact parameter of \mbox{$(18 \pm 13)$~$\mu$m}. It is incompatible with originating from the secondary vertex,
 the impact parameter being $(96 \pm 12)$~$\mu$m.
\end{itemize}

A scanning procedure \cite{largeangle} with an extended angular
acceptance (up to $\tan \theta =3.5$) did not reveal any
additional large-angle primary track that could be left by a muon or an electron.

In a dedicated search in the 19 films downstream of the primary vertex and in the 10 most upstream films of the downstream brick (more than
5~$X_0$ in total), no further $\gamma$-ray shower within a slope
acceptance of $\tan \theta<1$ and an energy above 
500~MeV could be found.  The single $\gamma$-ray detected at the
primary vertex can be interpreted as coming from the decay of a
$\pi^0$ with the other $\gamma$ being undetected.

\begin{figure}
\centering
\begin{center}
\includegraphics[width=7.2cm,height=7.2cm]{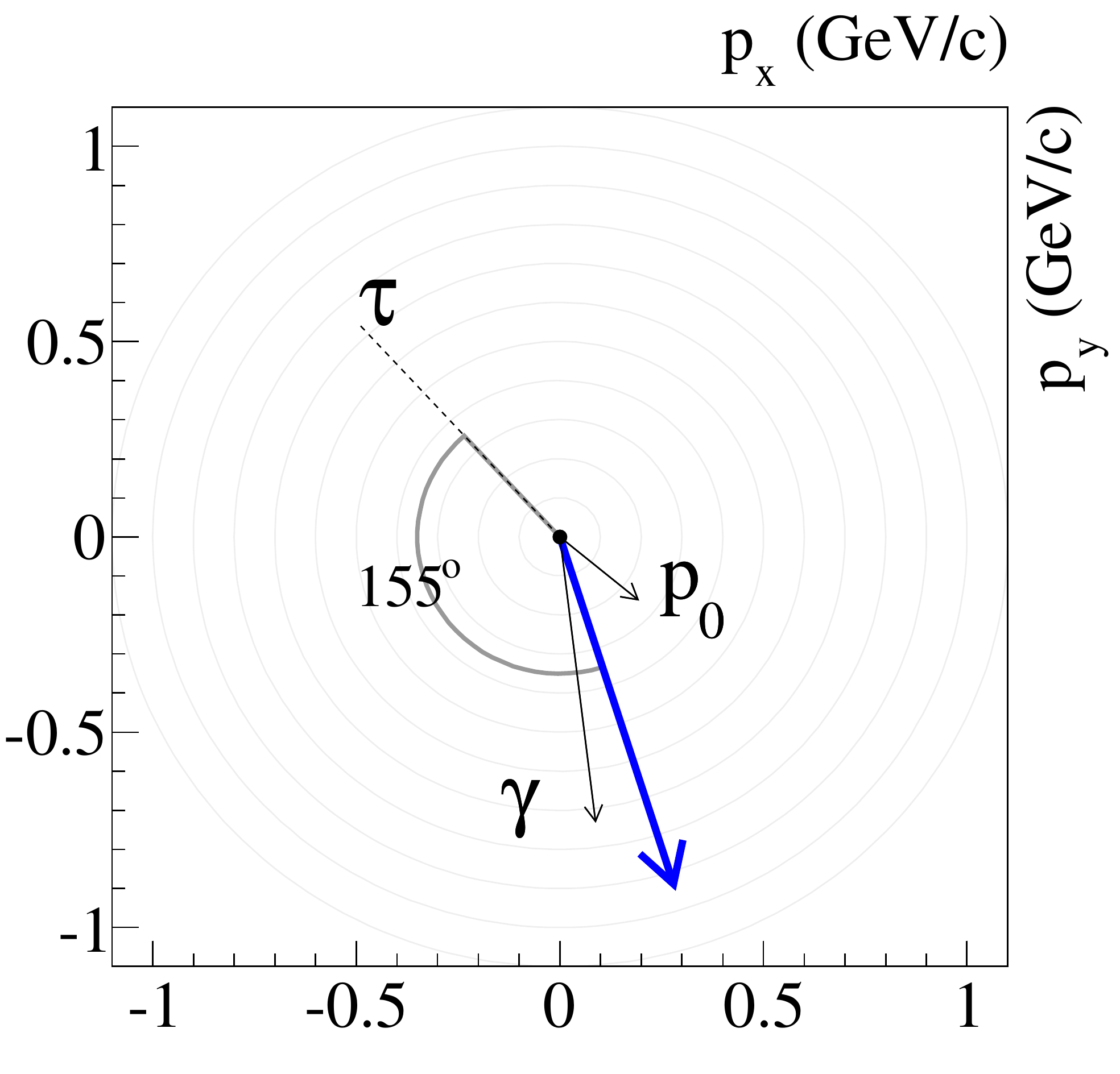}
\end{center}
\caption{The $\tau$ lepton direction (dashed line) and the momenta of the other primary
  particles ($p_0$ and $\gamma$-ray) in the plane transverse to the CNGS
  beam. The blue arrow is the sum of the transverse momenta of
  the $p_0$ track and of the $\gamma$-ray.
\label{fig:phi}}
\end{figure}

In the plane transverse to the beam direction, the angle between the
$\tau$ candidate direction and the sum of the transverse momenta of
the other primary particles ($p_0$ and $\gamma$) is
\mbox{$\Delta\phi_{\tau H} = (155 \pm 15)^\circ$} (see
Fig.~\ref{fig:phi}).  The transverse momentum at the secondary vertex
($p_{T}^{2ry}$) amounts to $(690 \pm 50)$~MeV/$c$.  The scalar sum of the momenta of all the particles is \mbox{$p_{sum}=(6.8 ^{+0.9}_{-0.6})$~GeV/$c$}.

\begin{figure}
\centering
\begin{center}
\includegraphics[width=10cm,height=12cm]{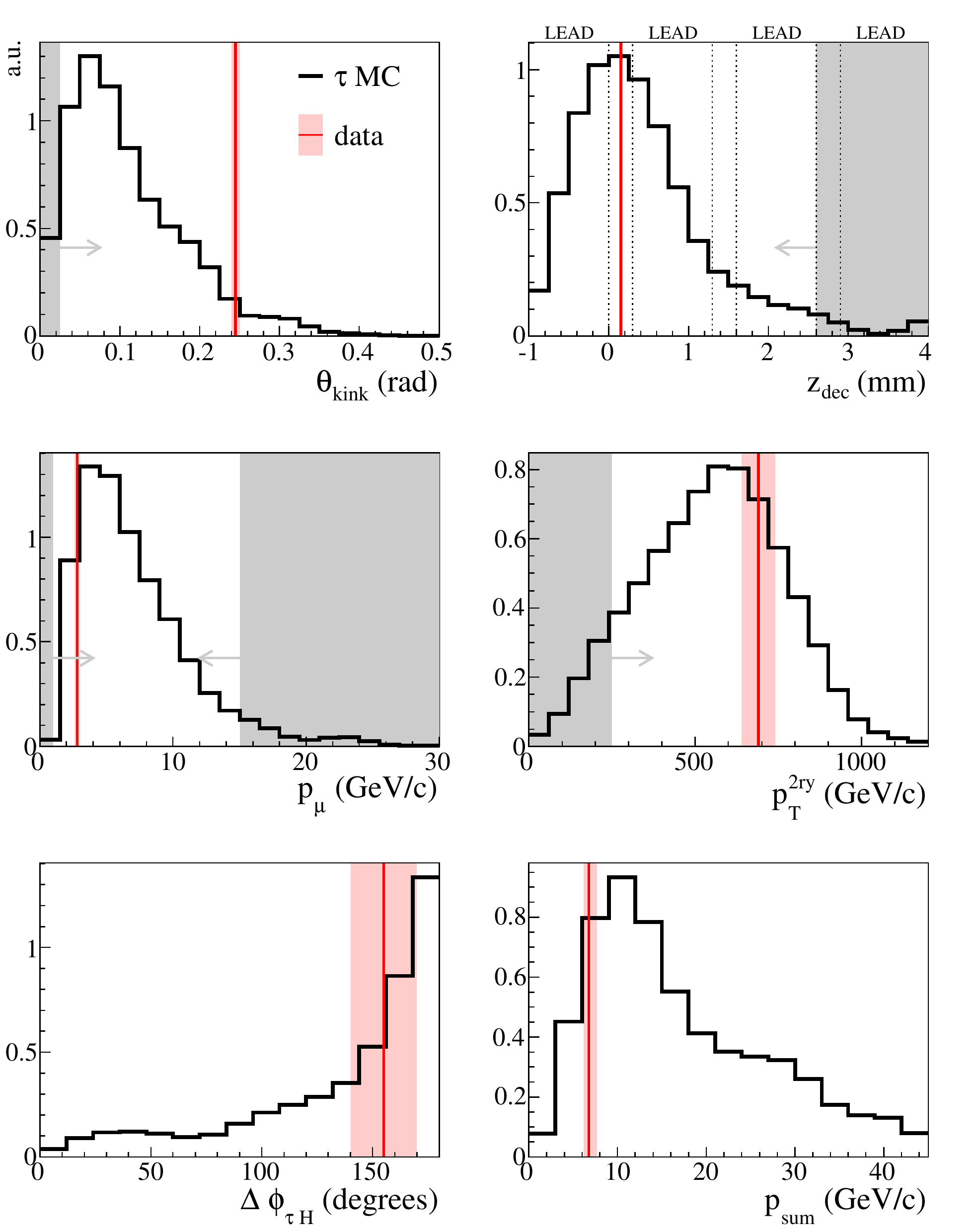}
\end{center}
\caption{Monte Carlo distribution of the reconstructed kinematical variables (see
  the text) for the $\tau\to \mu$
  decay channel.
Red  lines show the measured values and red bands their
  uncertainty. Grey areas cover the regions excluded by the selection
  cuts.}
\label{fig:fig9}
\end{figure}

In Tab. \ref{tab:tab2} the values of the kinematical
variables for this event are reported along with the predefined selection
criteria \cite{secondtau} 
for the $\tau \to \mu$ channel.
Besides satisfying all the selections, the variables are well within the domain of the
expected signal, see Fig.~\ref{fig:fig9}. 

\begin{table}
\centering
\begin{tabular}{ccc}
\hline
Variable & Selection ($\tau\to\mu$)& Measurement\\
\hline
$\theta_{kink}$ (mrad)& $>$ 20  &$245 \pm 5$\\
$z_{dec}$ ($\mu$m) &$<$ 2600 & $151 \pm 10$\\
$p_\mu$ (GeV/$c$) & $[1,~15]$ & $2.8 \pm 0.2$\\
$p_{T}^{2ry}$ (MeV/$c$) & $>$ 250 & $690 \pm 50$\\
\hline
\end{tabular}
\caption{Selection criteria for $\nu_\tau$ candidate events in the
  $\tau\to\mu$ decay channel along with the values measured
for the candidate event. Variables are defined in the text\label{tab:tab2}. }
\end{table}

\section{Expectations and statistical significance}
The method used for the estimation of signal and background was 
recently discussed \cite{secondtau}. With respect to those results, here is
also taken into account the extension of the analysed sample to the 1$\mu$
events of the 2011 and 2012 runs with $p_\mu<15$~GeV/$c$ (presently completed at 56\%).  

The total sample of analysed events is 5272 giving an expected $\nu_\tau$ signal in all decay channels of 1.7 events ($\Delta m^2_{23}=2.32 \times 10^{-3}$~eV$^2$ 
 and $\sin^22\theta_{23}=1$), out
of which 0.54 in the $\tau\to\mu$ decay channel.

For the background evaluation, the full sample of $1\mu$ events was conservatively
accounted for, although it is not yet completely analysed. The additional background only affects the $\tau \to \mu$ decay
channel which increases from $0.011 \pm 0.05$ (\cite{secondtau}) to
$0.021 \pm 0.010$ events, accordingly rising the total background to
$0.184 \pm 0.025$ events.  The background in the $\tau\to\mu$ channel
is dominated by the contribution of large-angle muon
scattering in lead (about 80\%) followed by charmed particle decays
(20\%), the background from hadronic interactions with a fake-muon 
being negligible \cite{secondtau}.

Accounting for the fact that the signal-to-background ratio is
different for each decay channel, the following method was
adopted: four Poissonian random integers  
are extracted, one for each decay channel in the background-only
hypothesis.  The $p$-values of the single channels (obtained as the integral of the Poisson distribution for values larger or equal to the observed number of candidates)
are combined into an estimator \mbox{$p^\star=p_\mu p_e p_h p_{3h}$}. 
By counting the fraction of extractions for which \mbox{$p^\star\leq
p^\star({\rm{observed}}$)}, the procedure allows excluding the absence of a $\nu_\mu \to \nu_\tau$ oscillation signal with a significance of {\color{black}{$3.4~\sigma$}} ($p$-value $= 2.9\times10^{-4}$).

Finally, it should be noted that the new candidate is in a region of
the parameter space that is free from background: 
the $\tau$ decay occurs in a low-density and low-$Z$ material (the plastic base) and with a transverse momentum at the secondary vertex of 690~MeV/$c$,
thus highly disfavouring the hypothesis of a large-angle muon
scattering.

\section{Conclusions}
The results of a $\nu_\tau$ appearance analysis on an extended sub-sample of the neutrino interactions collected by the
OPERA experiment in the CNGS run years 2008 to 2012 are reported.  
A $\nu_\tau$ candidate event
in the $\tau^-\to \mu^-$ decay channel was observed.  
A measurement of the negative charge of the $\tau$ lepton candidate, consistent with what is expected for the \mbox{$\nu_\mu \to \nu_\tau$} oscillation, has been performed for the first time.
With the present statistics and the observation of three $\nu_\tau$ candidates, 
the absence of a signal from $\nu_\mu\to\nu_\tau$ oscillations is excluded at 
\mbox{{\color{black}{3.4~$\sigma$}}}.

\section*{Acknowlegements}
We thank CERN for the successful operation of the CNGS facility and INFN for 
the continuous support given to the experiment through its LNGS laboratory. 
We acknowledge funding from our national agencies. Fonds de la Recherche 
Scientifique-FNRS and Institut InterUniversitaire des Sciences Nucl\'eaires for 
Belgium, MoSES for Croatia, CNRS and IN2P3 for France, BMBF for Germany, 
INFN for Italy, JSPS, MEXT, QFPU - Global COE programme of Nagoya University)
and Promotion and Mutual Aid Corporation for Private Schools of Japan for Japan, 
SNF, the University of Bern and ETH Zurich for Switzerland, the Russian Foundation 
for Basic Research (grant 12-02-12142 ofim), the Programs of the Presidium of the 
Russian Academy of Sciences (Neutrino physics and Experimental and theoretical 
researches of fundamental interactions), and the Ministry of Education and Science 
of the Russian Federation for Russia, the National Research Foundation of Korea 
Grant No. 2011-0029457 and TUBITAK for Korea, the Scientific and Technological 
Research Council of Turkey, for Turkey. We thank the IN2P3 Computing Centre 
(CC-IN2P3) for providing computing resources.

\bibliographystyle{prsty}
\bibliography{bib}

\end{document}